\documentclass[aps,prl,twocolumn,amsmath,amssymb,superscriptaddress]{revtex4-1}
\usepackage{amsmath,amssymb,bm}

\usepackage{graphicx}

\begin{document}
\title{$Z_N$ Berry Phases in Symmetry Protected Topological Phases}

\date{\today}

\author{Toshikaze Kariyado}\email{kariyado.toshikaze@nims.go.jp}
\affiliation{International Center for Materials Nanoarchitectonics
(WPI-MANA), National Institute for Materials Science, Tsukuba, 305-0044,
Japan}
\author{Takahiro Morimoto}
\affiliation{Department of Physics, University of California, Berkeley}
\author{Yasuhiro Hatsugai}
\affiliation{Division of Physics, University of Tsukuba}

  \begin{abstract}
    We show that the $Z_N$ Berry phase (Berry phase quantized into $2\pi/N$) provides a useful tool to characterize symmetry protected topological phases
    with correlation that can be directly computed through numerics of a relatively small system size.
   The $Z_N$ Berry phase is defined in a $N-1$ dimensional
   parameter space of local gauge twists, which we call ``synthetic
   Brillouin zone'', and an appropriate choice of an integration path consistent with the symmetry of the system ensures exact quantization of the Berry phase. We demonstrate the usefulness of the $Z_N$ Berry phase by studying two 1D models of bosons, SU(3) and SU(4) AKLT models, where topological phase transitions
   are captured by $Z_3$ and $Z_4$ Berry phases, respectively. 
We find that the exact quantization of the $Z_N$ Berry phase at the topological transitions arises from a gapless band structure (e.g., Dirac cones or nodal lines) in the synthetic Brillouin
   zone.
  \end{abstract}

  \maketitle

  In the past decades, topology has come
  to the fore of the condensed matter research and it has been realized
  that it serves as a guiding principle to explore novel phases of
  matter without relying on the symmetry breaking
  \cite{PhysRevB.40.7387}. Meanwhile, symmetry still plays an important role
  in an interplay with topology. For example, topological phases of noninteracting
  fermions have been classified according to the generic internal
  symmetries, i.e., time-reversal, particle-hole, and
  chiral symmetries
  \cite{PhysRevB.40.7387,PhysRevB.78.195125,DOI:10.1063/1.3149495,1367-2630-12-6-065004,1367-2630-12-6-065010}.
The topological classification of noninteracting fermions has been further extended by incorporating crystal
  symmetries
  \cite{PhysRevLett.106.106802,PhysRevB.88.075142,PhysRevB.88.125129,PhysRevB.90.165114,1367-2630-12-6-065004,Po:2017aa,Bradlyn:2017aa}.
On the other hand, characterization of topological phases becomes a more difficult problem for systems of strongly interacting particles \cite{PhysRevB.81.134509}. There have been active studies on
  classification and characterization
  of symmetry protected topological (SPT) phases that are supported with strong correlation effects
  \cite{PhysRevB.40.7387,doi:10.1143/JPSJ.75.123601,0953-8984-19-14-145209,PhysRevB.78.054431,PhysRevB.82.155138,PhysRevB.85.075125,PhysRevB.86.125119,PhysRevB.90.245120,PhysRevB.92.081304,PhysRevB.92.125104}.
However, the characterization of SPT phases
  for a given Hamiltonian remains a highly nontrivial
  problem. In particular, a concise way to characterize them through  fairly cheep numerics has been desired.

In characterizing SPT phases, the notion of adiabatic
continuation plays an essential role \cite{doi:10.1143/JPSJ.75.123601,0953-8984-19-14-145209,PhysRevB.78.054431,PhysRevB.90.085132}.
By adiabatically continuing a given system into a simple reference system,
the topological character in the original system is easily diagnosed by studying the simple reference system. 
For example, a system that can be
adiabatically decomposed into a set of the elementary units in the
system (an atomic insulator in the case of free fermions) is identified
as a trivial phase. In contrast, the requirement for keeping a finite
gap and the symmetry of the system sometimes excludes
possibility of ``atomic insulators'', and leaves a set of finite-size
\textit{entangled} clusters
\cite{doi:10.1143/JPSJ.75.123601,PhysRevB.82.155138,refId0}, which
indicates that the state is in an SPT phase. A representative
example is Haldane phase in a spin-1 Heisenberg chain 
\cite{HALDANE1983464,PhysRevLett.50.1153,0953-8984-19-14-145209,PhysRevB.45.2207,PhysRevB.85.075125},
where the entangled clusters are \textit{intersite} singlets of emergent
spin-1/2 degrees of freedom.

\begin{figure}[tbp]
 \includegraphics{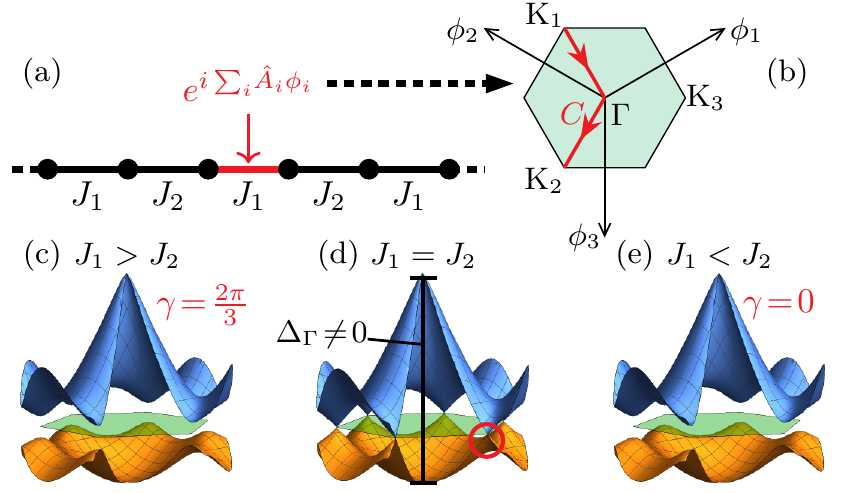}
 \caption{(Color online) (a) Schematic picture of the biquadratic model and the local gauge  twist. The bonds $J_i$ represent the biquadratic interactions. (b)
 ``Synthetic'' Brillouin zone and the integration path $C$ leading to
 the Berry phase quantization. (c-e) The energy spectra for the ground
 state and the first excited state on the Brillouin zone. The gap at
 $\Gamma$ point is always nonzero due to the finite size effect. At the
 phase transition, the gap closes at K and K' points forming Dirac
 cones.}\label{fig:concept}
\end{figure}

In the search of adiabatic continuation into the embedded entangled clusters,
it is useful to study Berry phase defined through the local
gauge twist [as schematically illustrated in Fig.~\ref{fig:concept}(a)]
\cite{doi:10.1143/JPSJ.75.123601,PhysRevB.78.054431,PhysRevB.88.184418,PhysRevB.90.085132,PhysRevB.91.214410}. 
Since the Berry phase can be quantized by symmetry in some cases, it provides a
conserved quantity in the process of the adiabatic continuation that encodes the topological nature of the system.
The Berry phase for the entangled cluster is easily obtained in the simple
reference system, and gives a characterization for the
original system. For instance, the
spin-1/2 singlet in Haldane phase in spin chain is characterized by
Berry phase $\pi$ \cite{doi:10.1143/JPSJ.75.123601}.
While the analysis based on Berry phase is useful in characterizing SPT phases, 
studies of correlated systems  so far have mainly focused on those phases characterized by Berry phase $\pi$. 

In this paper, we generalize the characterization of SPT phases
with correlation based on Berry phase by using fractionally quantized
Berry phase $2\pi/N$ ($Z_N$ Berry phase), 
and propose that such $Z_N$ Berry phase provides a useful tool to
diagnose general topological phases of interacting particles \cite{refId0}.
We demonstrate that the $Z_N$ Berry phase is useful in characterizing
one-dimensional SPT phases classified by general $Z_N$ topological number.
Specifically, we extend the Berry phase analysis so that it can detect entangled clusters other than the conventional spin-1/2
singlets. 
We demonstrate that spin-1 singlets can be detected with the appropriate redefinition of
the Berry phase.
This can be applied to a bond
alternating spin-1 chain with biquadratic interaction (hereafter, called the
biquadratic model), which supports a $Z_3$ SPT phase.  
In this case, the phase transition is captured by the $Z_3$ Berry phase (0 or $2\pi/3$), instead of the conventional one $\pi(=2\pi/2)$. 
We also show that an SU(4) symmetric spin chain supports a topological
phase with an SU(4) fully antisymmetrized state being the entangled cluster,
which can be diagnosed by $Z_4$ Berry phase. 
These generalizations of the Berry phase into fractional ones
 involve ``synthetic''
Brillouin zone (BZ) [see Fig.~\ref{fig:concept}(b)] that parameterizes local gauge twists for a particular bonds.
When there exist $N$ kinds of local gauge twists 
[$N=3$ for the SU(3) chain and $N=4$ for the SU(4) chain], 
such synthetic BZ is given by a $N-1$ dimensional space. 
(Note that the system itself is one-dimensional.)
We find that the phase transition is governed by a gapless structure appears in the effective band structure in the synthetic Brillouin
zone such as Dirac cones shown in
Fig.~\ref{fig:concept}(d).
Thus the $Z_N$ Berry phase analysis allows us to understand the topological phase transition in the many-body system by using an analogy to that in free-fermion system.

Let us begin with the formulation of the Berry phase.
For simplicity, we focus on a one-dimensional periodic system
with Hamiltonian $H=\sum_{ij}H_{ij}$. For finite size systems (either open or periodic) that are studied by numerical calculations in practice, the Berry phase is defined in the following way. 
First, we pick up a term on a certain bond, $H_{nm}$, out of the terms in the Hamiltonian. Then, it is
replaced by $U_m(\phi)H_{nm}U^\dagger_m(\phi)$, 
where $U_m(\phi)=e^{i\hat{A}\phi}$ (the local gauge twist) acts on the $m$th
site. While it looks like a unitary transformation, it actually is not, since the
operation is selectively acting on the chosen bond. Therefore, the eigenvalues
and eigenvectors change as
$|G_0\rangle\rightarrow |G_\phi\rangle$. Using the set of these wave functions, the Berry phase $\gamma$ is defined as
\begin{equation}
 i\gamma = \int_0^{2\pi}d\phi \langle G_\phi|\partial_\phi G_\phi\rangle.
\end{equation}
The choice of the gauge twist $\hat{A}$ is the most important part of this scheme. It should make
$U_m(\phi)$ periodic in $\phi$, and should properly capture the
underlying entangled cluster. 

\begin{table}[tb]
 \begin{center}
  \caption{Extra phase factors associated with each term of the
  biquadratic model induced by the dipolar and quadrupolar twist.}\label{table1}
  \begin{tabular}{|c|c|c|c|}
   \hline
   term &expression & dipolar & quadrupolar \\
   \hline
   1&$S^z_nS^z_nS^z_mS^z_m$& $1$ & $1$\\
   2&$S^z_nS^+_nS^z_mS^-_m/2$& $e^{i\phi}$ & $e^{-i\phi}$\\
   3&$S^z_nS^-_nS^z_mS^+_m/2$& $e^{-i\phi}$ & $e^{-i\phi}$\\
   4&$S^+_nS^z_nS^-_mS^z_m/2$& $e^{i\phi}$ & $e^{i\phi}$\\
   5&$S^+_nS^+_nS^-_mS^-_m/4$& $e^{2i\phi}$ & $1$\\
   6&$S^+_nS^-_nS^-_mS^+_m/4$& $1$ & $1$\\
   7&$S^-_nS^z_nS^+_mS^z_m/2$& $e^{-i\phi}$ & $e^{i\phi}$\\
   8&$S^-_nS^+_nS^+_mS^-_m/4$& $1$ & $1$\\
   9&$S^-_nS^-_nS^+_mS^+_m/4$& $e^{-2i\phi}$ & $1$\\
   \hline
  \end{tabular}
 \end{center}
\end{table}
In the previous studies of spin systems, $\hat{A}=S-\hat{S}_z$ has been the
standard choice
\cite{doi:10.1143/JPSJ.75.123601,PhysRevB.78.054431,PhysRevB.88.184418},
which is suitable for detecting a
spin-1/2 singlet. 
In this case, some symmetries constrain the Berry phase $\gamma$ 
to quantize into $0$ or $\pi$, where
$\gamma=\pi$ signals the existence of a spin-1/2 singlet at the chosen
bond. This is indeed the case for the Haldane
phase in the spin-1 Heisenberg chain, which is a representative SPT
phase. The topological nature of the Haldane phase is captured by the valence bond
solid picture, where pairs of spin-1/2 obtained from fractionalization of the original spin-1 form \textit{intersite} spin-1/2
singlets \cite{PhysRevLett.59.799}. 
The Berry phase quantizes into $\gamma=\pi$ in the
Haldane phase,
while it quantized into $\gamma=0$ in the topologically trivial large-D phase where fractional spin-1/2's form \textit{intrasite} spin-1/2
singlets. 
Such quantization of the Berry phase (into $0$ or $\pi$) allows us to observe the sharp transition between the
Haldane and the large-D phases even for a chain of a relatively small
number of sites.
This observation can be generalized to the case of $Z_N$ Berry phase. Namely,
the quantization of the general $Z_N$ Berry phase indicates a sharp
phase transition even for a small size system (without extrapolation to
the thermodynamic limit) that can be studied in practical numerical
calculations.

\begin{figure}[tbp]
 \includegraphics{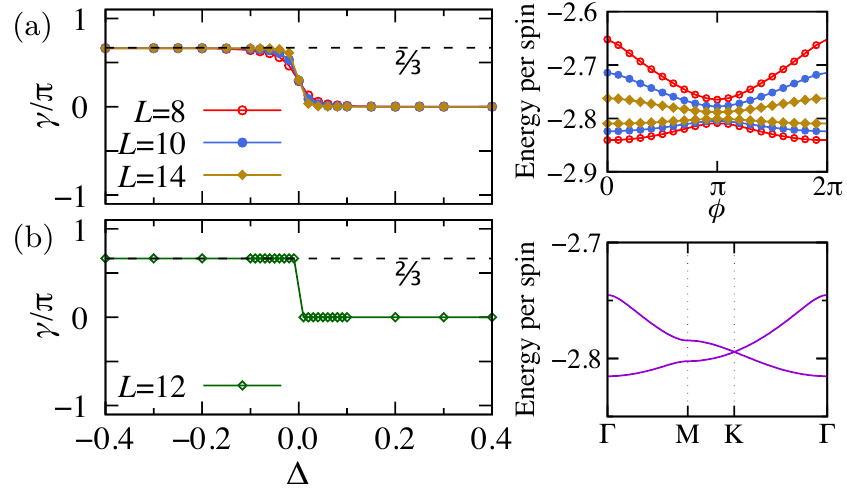}
 \caption{(Color online) (a) The Berry phase with $e^{i\hat{A}_3\phi}$ for several
 system size ($L$ denotes the number of the spins). The right panel shows the
 energy spectra as functions of $\phi$ at
 $\Delta=0$. (b) The Berry
 phase with
 $e^{i\sum_i\hat{A}_i\phi_i}$ using $C$ in Fig.~\ref{fig:concept}(b)
 as an integration path. The right panel shows the energy spectrum along the
 high symmetry lines in the synthetic Brillouin zone. $L$ is set to
 $12$. The energies are in the unit of $J$.} \label{fig:biq}
\end{figure}

Next, we study a case where the entangled cluster is not a conventional
spin-1/2 singlet. To this end, we consider a spin-1 chain with
bond-alternating biquadratic interaction \cite{PhysRevB.40.4621}, which is described by the Hamiltonian,
\begin{equation}
 \hat{H}=-J_1\sum_i(\hat{\bm{S}}_{2i}\cdot\hat{\bm{S}}_{2i+1})^2
  -J_2\sum_i(\hat{\bm{S}}_{2i+1}\cdot\hat{\bm{S}}_{2i+2})^2. \label{eq:biqH}
\end{equation}
It is known that this model supports the SU(3) AKLT state \cite{PhysRevB.90.235111}. In the
language of the SU(3) AKLT state, the elementary object is a quark (and
antiquark) and the entangled cluster characterizing SPT phase is a meson
(specifically, $\eta$-meson). In the language of the spin-1
biquadratic model, the entangled cluster is mapped to a spin-1 ``singlet''
(two-spin state with zero total angular momentum).
By writing $J_{1,2}$ as
$J_1=J+\Delta$ and $J_2=J-\Delta$, the parameter $\Delta$ controls 
how the entangled cluster are formed. Therefore, once we fix the
position of the gauge-twisted bond, the transition has to be
observed by changing $\Delta$. 
However, the standard choice of $\hat{A}=S-\hat{S}_z$ is inadequate
for detecting the spin-1 singlet.
Instead, we use the twist operator $\hat{A}=\hat{A}_3\equiv 1-\hat{S}_z^2$. 
If the bond with the biquadratic interaction is twisted by
$e^{i\hat{A}_3\phi}$, each term acquires
the phase as summarized in Table~\ref{table1}. 
For comparison, we list the phase factors
acquired in the conventional twist with $e^{i(S-\hat{S}_z)\phi}$.
 Since
$\hat{S}_z$ ($\hat{S}_z^2$) is the part of the dipole (quadrupole)
moment, we call $e^{i(S-\hat{S}_z)\phi}$ ($e^{i\hat{A}_3\phi}$) dipolar
(quadrupolar) twist. An important
feature that we can see from the phase factors in Table~\ref{table1} is
their symmetry. If they are
symmetric with respect to the combined operation of the complex
conjugation and $n\leftrightarrow m$, the Berry phase should be
quantized into $0$ or $\pi$ \cite{doi:10.1143/JPSJ.75.123601}. 
Indeed the dipolar twist obeys this symmetry 
(e.g., the operation on the term 2 results in the term 3, leaving the
term unchanged in total). On the other hand, the quadrupolar twist breaks
this symmetry, and hence, it does not show $Z_2$ quantization, but may
quantize into other fractions of $2\pi$.

The numerically obtained Berry phase as a function of $\Delta$ is
summarized in Fig.~\ref{fig:biq}. We identify two phases, which are
characterized by $\gamma=0$ for $\Delta>0$ and $\gamma=2\pi/3$ for
$\Delta<0$. The embedded spin-1 singlet exists on the twisted bond
if $\gamma=2\pi/3$, since 
an isolated singlet with a $\hat{A}_3$-twist is described by the wave function,
$|\psi_\phi\rangle=(|{+1},{-1}\rangle-e^{i\phi}|0,0\rangle+|{-1},{+1}\rangle)/\sqrt{3}$ 
(by using a representation for the state of two spins, $|s^{z}_1,s^{z}_2\rangle=|s^{z}_1\rangle\otimes|s^{z}_2\rangle$
with $\hat{S}_{zi}|s^{z}_i\rangle=s^{z}_i|s^z_{i}\rangle$), and the second term $e^{i\phi}|0,0\rangle$ contributes to the Berry phase by $2\pi/3$.
Note that with the dipolar
twist, the Berry phase is zero regardless of the sign of $\Delta$, which
means that the phase transition in Fig.~\ref{fig:biq} is captured only
with our new method. The system size dependence in Fig.~\ref{fig:biq}(a)
suggests that the transition gets sharper as we approach the
thermodynamic limit. However, the quantization of
the Berry phase is not perfect. Thus it does not ensure the advantage of using the
Berry phase, i.e., quantization even for a relatively small size system. 

Fortunately, we have a remedy to this deviation from perfect quantization. The reason
why it does not show quantization is that the
symmetry of the system is not fully appreciated.
The key symmetry of Eq.~\eqref{eq:biqH} is
the spin rotational symmetry, in particular, the symmetry under the interchange of
$x$, $y$, and $z$-directions in the spin space.
(This corresponds to the 
interchange of three flavors of quarks which forms a $Z_3$ subgroup of the SU(3) symmetry
\cite{PhysRevB.90.235111}.)
Accordingly, our formulation of Berry phase can be symmetrized by considering the other twist operators
$\hat{A}_1=1-\hat{S}_x^2$ and $\hat{A}_2=1-\hat{S}_y^2$ in addition to
$\hat{A}_3$, and we define the generalized local gauge twist as
$\exp[i\sum_i\hat{A}_i\phi_i]$ with three parameters
$\phi_{1,2,3}$. Because of $\hat{A}_1+\hat{A}_2+\hat{A}_3=\hat{1}$,
only two of three parameters are independent, namely, a twist by 
$e^{i\hat{A}_3\phi}$ has the same effect as a twist by
$e^{-i(\hat{A}_1+\hat{A}_2)\phi}$ since $e^{i\hat{1}\phi}$ is trivial. This means that the generalized local gauge twist is 
defined on the two-dimensional periodic parameter space, which we call
``synthetic Brillouin zone'', with the hexagonal symmetry
as shown in Fig.~\ref{fig:concept}(b). 
In terms of the synthetic BZ, we can see that the Berry phase defined for a straight line along the
$\phi_3$ axis in the synthetic BZ leads to deviation from the quantization [Fig.~\ref{fig:biq}(a)]. 
Instead, we now consider the path $C$
(K$_1$-$\Gamma$-K$_2$) in Fig.~\ref{fig:concept}(b) which is more symmetric in the synthetic BZ. 
Figure~\ref{fig:biq}(b) shows the Berry phase obtained with the
path $C$. In this case, the Berry phase shows an exact quantization into $0$ and
$2\pi/3$, leading to the sharp transition. The origin of the quantization is
understood by considering the Berry phases defined with three different
paths, $\gamma_1$ with K$_1$-$\Gamma$-K$_2$, $\gamma_2$ with 
K$_2$-$\Gamma$-K$_3$, and $\gamma_3$ with K$_3$-$\Gamma$-K$_1$. By the
three-fold rotational symmetry in the synthetic BZ, we obtain $\gamma_1=\gamma_2=\gamma_3$. At the same
time, if the three paths are combined, they
result in a trivial path, giving us 
$\sum_i\gamma_i=0$ (mod $2\pi$). The consequence of this symmetry consideration is that the Berry phase $\gamma_i$ should quantize into $2\pi/3$ \cite{refId0}. 

The introduction of the synthetic BZ reveals another
notable aspect of the transition, i.e., an emergent gapless structure in the effective band structure. Generally speaking, 
quantization of the Berry phase indicates a jump in
the value of $\gamma$ at the phase transition, and such a jump
requires a singularity in the wave function which is associated with gap closing. In this case, the energy gap above the
ground state should close somewhere on the integration path. 
Conversely, no sharp transition is expected when the gap remains 
finite over the entire integration path. The right panel of Fig.~\ref{fig:biq}(a) plots the energy spectrum as
a function of $\phi$ for the $e^{i\hat{A}_3\phi}$ twist at $\Delta=0$,
which shows the absence of any gap closure. This accounts for the smooth
change of $\gamma$ at $\Delta$ in Fig.~\ref{fig:biq}(a). 
In contrast, we indeed have a gap
closing point on the path $C$ at $\Delta=0$. More specifically, the
gapless points are found at K and K' points in the synthetic Brillouin
zone. [See Figs.~\ref{fig:concept}(c-e) and the right panel of
Fig.~\ref{fig:biq}(b).] Interestingly, the energy spectrum at $\Delta=0$
shows Dirac cones, in a similar way to the band structure of graphene. This reminds
us the fact that the topological transition in free fermion systems is
often associated with a gapless band structure such as Dirac cones.
In an analogy, the topological phase transition in our model, although it is a correlated
one-dimensional model, is associated with the Dirac cones that appear in the ``synthetic''
Brillouin zone. Note that the gap at $\Gamma$ point, representing the
state without any twist, is always finite including the case with
$\Delta=0$. 
In passing, we note that it is known that
the ground state is doubly degenerate in the thermodynamic limit for $\Delta=0$ \cite{PhysRevB.40.4621}. 
This means that the ``band structures'' in
Figs.~\ref{fig:concept}(c-e) collapse in the infinite size
limit, and the jump in $\gamma$ gets sharper with $L\to \infty$ in any case. However, as we have stressed earlier, the
advantage of the quantized Berry phase lies in the usefulness in
the finite size calculation of a relatively small system size. 

\begin{figure}[tbp]
 \includegraphics{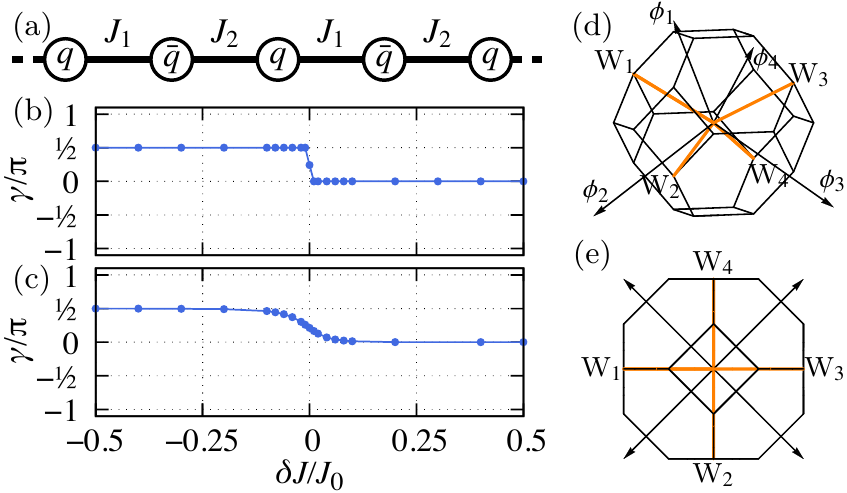}
 \caption{(Color online) (a) Schematic picture of the SU(4) model. $q$ and $\bar{q}$
 denote the fundamental representation and the conjugate representation,
 respectively. (b) The Berry phase obtained with the path
 W$_1$-$\Gamma$-W$_2$. (c) The Berry phase obtained using the straight
 integration path along $\phi_1$ axis. (d,e) Synthetic Brillouin zone
 and the integration path.} \label{fig:su4_Berry}
\end{figure}
Next we show the usefulness of the $Z_N$ Berry phase by applying it to
another example of 1D SPT phases. We consider an
SU(4) symmetric Hamiltonian \cite{PhysRevLett.54.966},
\begin{equation}
 H=-\sum_{i}\sum_{a=1}^{15}\bigl[J_1\Lambda_a(2i)\bar{\Lambda}_a(2i+1)+J_2\bar{\Lambda}_a(2i+1)\Lambda_a(2i+2)\bigr].
\end{equation}
Here, the fundamental representations of SU(4) and its
conjugate representations are assigned on the $(2i)$th sites and $(2i+1)$th sites, respectively [see Fig.~\ref{fig:su4_Berry}(a)].
The explicit form
of the $\Lambda_a$ is found in Ref.~\onlinecite{su4}. For convenience,
we parameterize $J_{1,2}$ as $J_1=J_0+\delta J$ and
$J_2=J_0-\delta J$. With the appropriate parameter choice, the ground
state of this Hamiltonian becomes to share the majority of properties
with the SU(4) AKLT state \cite{PhysRevB.90.235111}. 
In this case, the
entangled cluster is the completely antisymmetrized state formed
by a pair of the fundamental and its conjugate representations
(which is analogous to the $\eta$-meson in the SU(3) case). 
In a similar manner to the case of $\Delta$ for the biquadratic model, $\delta J$ controls how the entangled clusters are formed, and the phase transition takes place by changing $\delta J$. For the
detection of the pattern of entangled clusters, we adopt
$U(\bm{\phi})=\exp[i\sum_{n=1}^4\check{A}_n\phi_n]$ as a gauge twist,
where $(\check{A}_n)_{ij}=\delta_{ij}\delta_{in}$. By using $\sum_n\check{A}_n=\hat{1}$, we notice that a twist $e^{i\check{A}_4\phi}$ is
essentially equivalent to a twist
$e^{-i(\check{A}_1+\check{A}_2+\check{A}_3)\phi}$, and consequently, the
local gauge twist is defined on the three-dimensional synthetic
BZ with the symmetry of the fcc BZ. 

\begin{figure}[tbp]
 \includegraphics{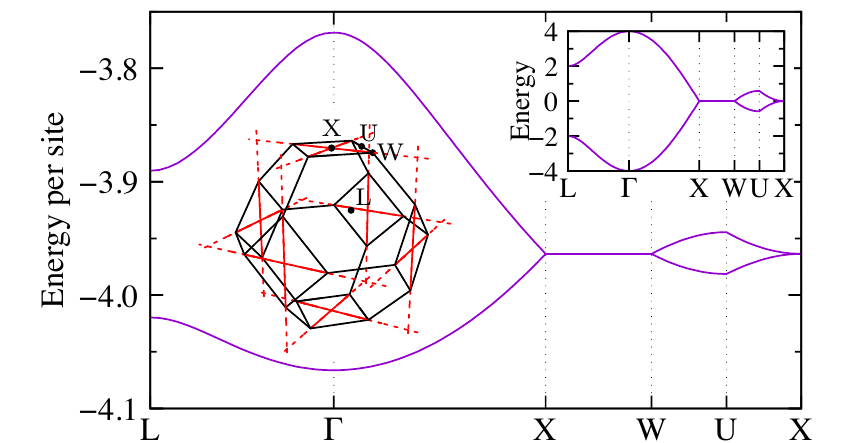}
 \caption{(Color online) Energy spectrum (in the unit of $J_0$) on the
 symmetric lines in the synthetic
 Brillouin zone for $\delta{J}=0$. The upper right inset shows the band
 structure of the
 single orbital tight-binding model on the diamond lattice. The left
 inset shows the location of the nodal lines.}\label{fig:su4_band}
\end{figure}
The numerically obtained Berry phase is plotted in
Fig.~\ref{fig:su4_Berry}. 
Again, the exact quantization of the Berry phase is achieved for a symmetric integration path W$_1$-$\Gamma$-W$_2$ in the synthetic
BZ as shown in Figs.~\ref{fig:su4_Berry}(d) and
\ref{fig:su4_Berry}(e).
With this setup, the phase transition is captured by a jump from $\gamma=0$ to
$\gamma=\pi/2=(2\pi/4)$
\cite{refId0,MotoyamaTodo}. Similarly to the SU(3)
case, the symmetry protecting the quantization of $Z_4$ Berry phase is the invariance under the interchange of the four components of the fundamental
representation of SU(4).
When the straight
integration path along one of
the $\phi_i$ axis is used naively, 
the Berry phase is no longer
quantized and it does not show a sharp transition at $\delta{J}=0$
[Fig.~\ref{fig:su4_Berry}(c)].
The jump in the $Z_4$ Berry phase is again associated
with the gapless point on the integration path. In this case, the gap closes on the X-W line, i.e., nodal
lines appear in the three-dimensional Brillouin zone for $\delta{J}=0$ [see
Fig.~\ref{fig:su4_band}]. 
Interestingly, the energy spectrum resembles
the band structure for the single orbital tight-binding model on the
diamond lattice.

To summarize, we have demonstrated the usefulness of the
$Z_N$ Berry phase as a
topological invariant for SU(N) symmetric SPT phases. The key ingredient is
a suitable choice of $\hat{A}$ for the local gauge twist, and the
introduction of the synthetic Brillouin zone reflecting the symmetry of
the system. The topological transitions are captured by jumps in the
Berry phase, and the associated singularities (Dirac cones/nodal lines)
in the synthetic Brillouin zone. It would be an interesting future problem to explore the relationship between
Dirac cones/nodal lines found here and those in free fermion systems at the
topological transition, for example, in terms of the criticality.
Another promising direction would be an
extension of the $Z_N$ Berry phase to topological phases in higher
spatial dimensions. The major task in doing so will be finding proper
ways of applying the local gauge twist to ensure exact
quantization. Once they are found, it will provide a tractable way to
characterize general SPT phases based on the Hamiltonians explicitly.

\begin{acknowledgments}
 TK thanks the Supercomputer Center, the Institute for Solid State
 Physics, the University of Tokyo for the use of the facilities. The
 work is partially supported by a Grant-in-Aid for Scientific Research
 No.~17K14358 (TK), No.~17H06138 (TK, YH)
 and No.~16K13845 (YH). TM was supported by the Gordon and Betty Moore
 Foundation's EPiQS Initiative Theory Center Grant.
\end{acknowledgments}

\end{document}